\documentclass{cpbtex}

\begin{document}

\title{\bf Progress on band structure engineering of twisted bilayer and two-dimensional moir\'e heterostructures\thanks{W.Y. and M.A. thank B. Stadtm\"uller (University of Kaiserslautern) for stimulating discussions. This work received financial support from the German Science Foundation via the Collaborative Research Centers 173 ``Spin + X: spin in its collective environment'' (project A02). S.Z. acknowledges the support from National Natural Science Foundation of China (Grant No.~11725418), Ministry of Science and Technology of China (Grants No.~2016YFA0301004), Science Challenge Project (No.~TZ2016004), Beijing Advanced Innovation Center for Future Chip (ICFC), Tsinghua University Initiative Scientific Research Program.}}

\author{Wei Yao$^{1}$, Martin Aeschlimann$^{1}$, Shuyun Zhou$^{2}$\thanks{Corresponding author. E-mail: syzhou@mail.tsinghua.edu.cn}\\
$^{1}${\small Department of Physics and Research Center OPTIMAS, University of Kaiserslautern, 67663 Kaiserslautern, Germany}\\
$^{2}${\small State Key Laboratory of Low Dimensional Quantum Physics and Department of Physics,}\\{\small Tsinghua University, Beijing 100084, P. R. China}
}

\date{\small \today}
\maketitle

\begin{abstract}
Artificially constructed van der Waals heterostructures (vdWHs) provide an ideal platform for realizing emerging quantum phenomena in condensed matter physics. Two methods for building vdWHs have been developed: stacking two-dimensional (2D) materials into a bilayer structure with different lattice constants, or with different orientations. The interlayer coupling stemming from commensurate or incommensurate superlattice pattern plays an important role in vdWHs for modulating the band structures and generating new electronic states. In this article, we review a series of novel quantum states discovered in two model vdWH systems --- graphene/hexagonal boron nitride (hBN) hetero-bilayer and twisted bilayer graphene (tBLG), and discuss how the electronic structures are modified by such stacking and twisting. We also provide perspectives for future studies on hetero-bilayer materials, from which an expansion of 2D material phase library is expected.
\end{abstract}



\section{Introduction}
The birth of graphene has led to the golden age of two dimensional (2D) material research in the past one and a half decades \cite{1.Geim_G-scipaper,2.Geim_G-rev,3.Geim_rise-of-G,4.Novoselov_2D-film}. During this period, numerous exotic phenomena in various 2D materials have been experimentally discovered when approaching the single layer thickness limit, e.g. Dirac-type quasiparticles in graphene\cite{5.Novoselov_massless-gas,6.ZhangYB_G-QHE,7.ZhouSY_Dirac}, valley-selective circular dichroism (valleytronics) in transition metal dichalcogenides (TMDC)\cite{8.Heinz_valleytronics,9.Cui_valleytronics,10.Feng_valleytronics}, and so on. There are so many members in the 2D material family that almost all possible electronic states can be found \cite{11.2D_number}, such as normal or topological insulators\cite{12.hBN_rev,13.ZhangYi_Bi2Se3}, indirect- or direct-gap semiconductors\cite{14.ZhangYi_MoSe2}, conventional \cite{15.Xi_NbSe2} or high-temperature superconductors\cite{16.ZhangYB_monoBSCCO}, and intrinsic quantum spin Hall states\cite{17.Fu_QSH,18.Tang_mono-WTe2,19.WuSF_QSH}.

Thanks to the weak interlayer coupling of 2D materials via van der Waals (vdW) interaction, it is easy to isolate each single sheet from its bulk form by exfoliation\cite{4.Novoselov_2D-film} or thin film growth. The strategy to obtain novel quantum electronic states is to stack individual 2D sheets together to build an even richer playground - van der Waals heterostructures\cite{20.Geim_vdw,21.Novoselov_vdw}, which is a practical implement of the ``more is different'' principle\cite{22.Anderson_MID} in the 2D materials. The new quantum states usually emerge in van der Waals heterostructures with fascinating large-scale patterns resulting from the lattice mismatch between the constituted layers. According to whether there is a long-ranged period or not, the superlattice patterns can be classified into two types. One type called moir\'e pattern, has long-ranged period and thus forms a commensurate superlattice. The other type without long-ranged translation symmetry forms an incommensurate superlattice that in some cases resembles quasicrystalline pattern. These superlattice patterns can be obtained by differing lattice constants or lattice orientations between the individual layers in the vdWHs, following which two model vdWHs - graphene/hexagonal boron nitride (hBN) hetero-bilayer and twisted bilayer graphene (tBLG) - are then experimentally constructed (as shown in Fig.~1). A lot of exotic quantum electronic states have been discovered in these two prototypical graphene-based vdWHs. For those with commensurate superlattices (basically with a mismatch in the lattice constants or a small twisting angle), the periodic moir\'e potential induced by the long-ranged moir\'e pattern leads to  interaction between original graphene Dirac cones and replica Dirac cones, which further contributes to the emergence of second-generation Dirac cones (SDCs) in graphene/hBN heterostructures\cite{29.LeRoy_SDC,30.Geim_SDC,38.WEY_SDC} and flat bands in tBLG at so-called magic twisting angle\cite{70.MacDonald_PNAS2011,74.CaoY_MA1,77.CaoY_MA2}. For incommensurate superlattices using quasicrystalline 30\textdegree-tBLG as an example, the lack of periodicity gives rise to one novel kind of mirrored Dirac cones, showing distinct motion of Dirac fermions in a non-periodic potential\cite{106.Yao_30tBLG,107.Ahn_30tBLG}. The discoveries of new quantum phenomena in the field of vdWHs are not limited to those mentioned above, instead, they are still in active and rapid development.

\begin{center}
    \includegraphics{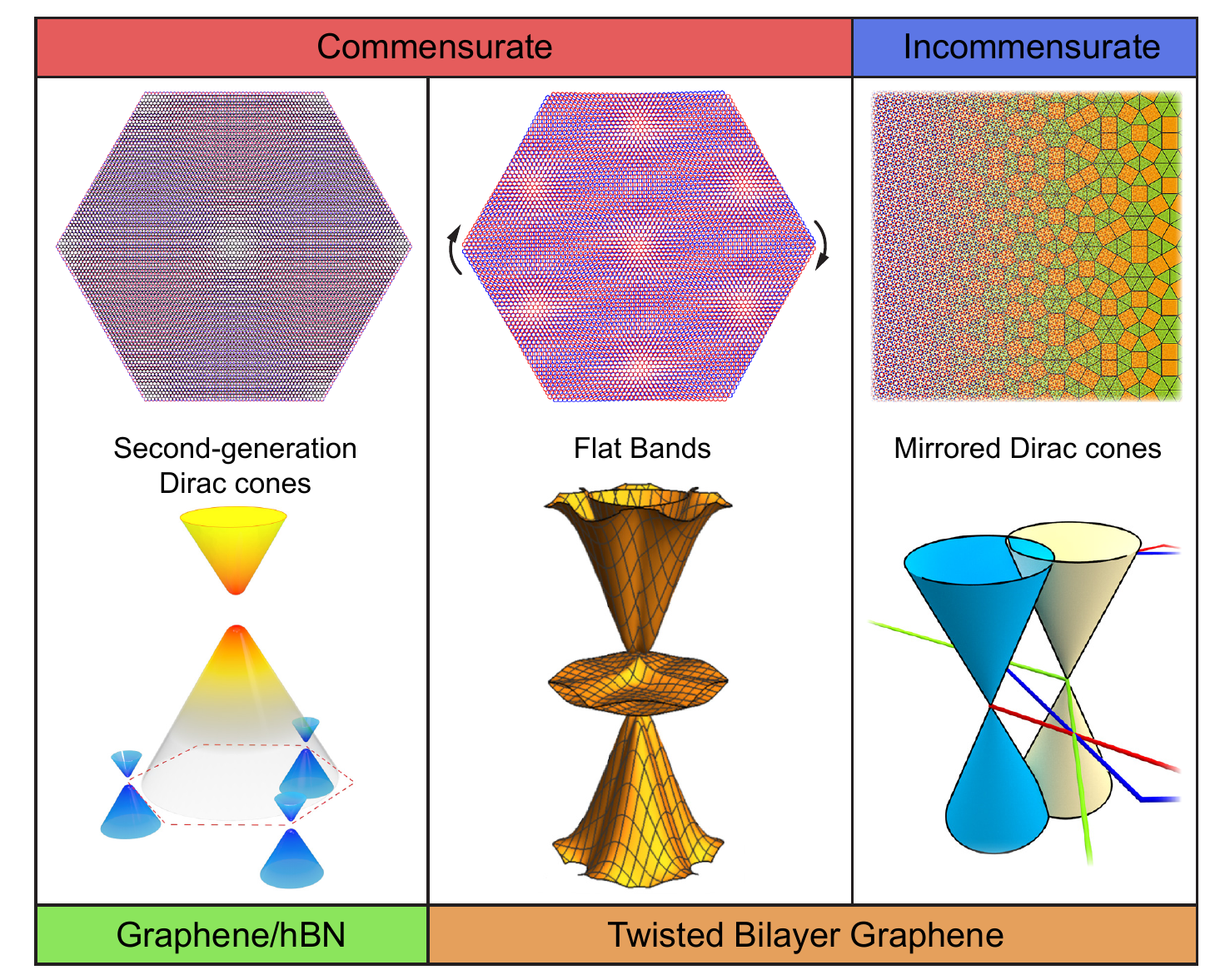}\\[5pt]  
    \parbox[c]{15.0cm}{\footnotesize{\bf Fig.~1.} (color online) Summary of commensurate and incommensurate van der Waals heterostructures with emerging novel quantum electronic states in graphene/hexagonal boron nitride and twisted bilayer graphene. Left: Superlattice pattern of graphene/hexagonal boron nitride heterostructure and corresponding second-generation Dirac cones (blue). Middle: lattice of twisted bilayer graphene at the magic angle and the corresponding flat bands which can induce strongly correlated phenomenon (Reprinted with permission from Ref.~\cite{31.flatband}. Copyright 2018, American Physical Society). Right: lattice of quasicrystalline 30\textdegree~twisted bilayer graphene and the corresponding mirrored Dirac cones. The graphene/hBN heterostructure and magic-angle tBLG are commensurate superlattices, while 30\textdegree-tBLG belongs to incommensurate superlattice.}
\end{center}

In this review, we discuss recent progress in the band structure engineering of vdWHs, mainly focusing on how the mismatched lattice structures modify the electronic properties in these model vdWHs. The band structure engineering and its impact on the electronic properties are discussed for commensurate and incommensurate structures respectively. Perspectives for future studies and applications are also provided in the last section.

\section{Energy bands in commensurate superlattices}
For commensurate superlattices, due to the conservation of the long-ranged translational symmetry, the electronic band structures are always strongly modified from the pristine ones by the moir\'e superlattice potential, e.g. the appearance of moir\'e replica bands and mini gap opening\cite{23.Kralj_G-Ir}. Such commensurate vdWHs have been most extensively investigated so far.

In general, for heterostructures formed by two single 2D-material sheets with  primitive lattice vectors of ${\bm a}_{1,2}$ and ${\bm b}_{1,2}$ respectively, the condition to satisfy commensurate superlattice is described as:
\begin{equation}
    n{\bm a}_1+m{\bm b}_1=n^{\prime}{\bm a}_2+m^{\prime}{\bm b}_2
\end{equation}
where $n$, $m$, $n^{\prime}$, $m^{\prime}$ are integers.

For heterostructures with different lattice constants and the same orientation, e.g. 0$^\circ$ aligned graphene/hBN heterostructure, the commensurate condition simply reduces to $n/n^{\prime}=|{\bm a}_2|/|{\bm a}_1|$, and the corresponding moir\'e period is $L=n|{\bm a}_1|=n^{\prime}|{\bm a}_2|$. Thus it is clear that the closer the lattice constants of the two layers ($|{\bm a}_1|$ and $|{\bm a}_2|$) such as graphene and hBN, the larger the integers $n$ and $n^{\prime}$, and the larger the moir\'e period $L$ as well.

For twisted bilayer structures like tBLG which consist of two layers with identical hexagonal lattice but different alignment (as shown in Fig.~2), the commensurate condition could be expressed by a series of discrete twisting angles\cite{24.CastroNeto_Continuum}: 
\begin{equation}
    \cos\theta_t=\frac{3p^2+3pq+q^2/2}{3p^2+3pq+q^2} \label{eq2}
\end{equation}
where $p$ and $q$ are two coprime positive integers. The moir\'e period can be derived as:
\begin{equation}
    L=a_0\sqrt{3p^2+3pq+q^2}
\end{equation}
where $a_0$ is the lattice constant for one single layer. Figure 2 illustrates the angular distribution for commensurate condition in one angular period (we choose $q=1$ here) with corresponding moir\'e period. One notable feature is that almost all commensurate structures are located near twisting angles of 0\textdegree~and 60\textdegree, which correspond to AA or AB stacked bilayer graphene. Therefore, most studies on commensurate tBLG basically focus on those with small twisting angles.

In this section, we review several novel quantum electronic states emerging in commensurate heterostructures, and discuss how the pristine band structures can be modified by the moir\'e superlattice potential and the resulting influence on the electronic properties.

\begin{center}
    \includegraphics{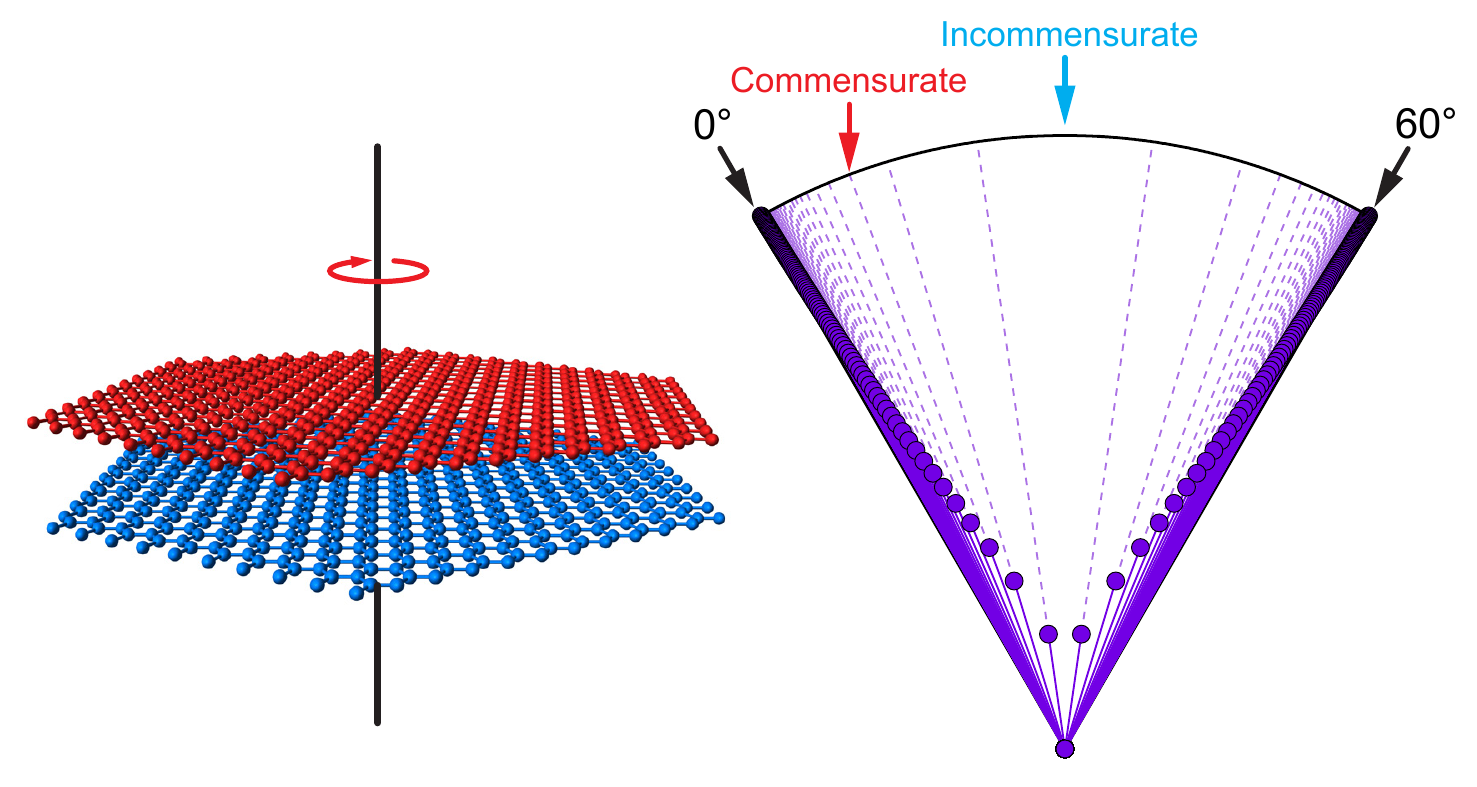}\\[5pt]  
    \parbox[c]{15.0cm}{\footnotesize{\bf Fig.~2.} (color online) Left: Twisted bilayer graphene is constructed by making relative rotation from an aligned bilayer graphene structure. Right: Angular distribution and corresponding superlattice period for commensurate condition in tBLG. The polar angles with respect to the 0\textdegree~edge in the fan shape define the twisting angles for commensurate tBLG. The distance from the center to the point defines the moir\'e superlattice period for each commensurate superlattice. Here we use the logarithmic scale.}
\end{center}

\subsection{Second-generation Dirac cones}

As a representative case for emerging electronic state in graphene vdWHs, second-generation Dirac cones (SDCs) are firstly discussed in the theoretical investigation of modified graphene band structures in the presence of a periodic potential\cite{25.CPark_pp_NP,26.CPark_pp_prl}. In general, due to the emergence of moir\'e replica Dirac cones and their interaction with the original graphene Dirac cones, a hybridization gap is usually expected at the original Dirac cones along the boundary of new moir\'e superlattice Brillouin zone (SBZ). However, as shown in Fig.~3(a), it has been predicted that there are gapless nodal points at the boundary of moir\'e SBZ due to the chiral nature of Dirac fermions in graphene\cite{26.CPark_pp_prl}. The band structures near these nodal points exhibit linear dispersions, hence they are called second-generation Dirac cones.

Experimental progress on SDCs is triggered by the successful realization of graphene/hBN vdWH by stacking graphene sheet on flat hBN  substrate\cite{27.Dean_hBN,28.LeRoy_hBN,28-1.Osterwalder_hBN}. The slight lattice mismatch of 0.2$\%$ between graphene and hBN naturally results in a moir\'e superlattice potential, thus being able to host SDCs. Experimental evidence of SDCs firstly appears in scanning tunneling spectroscopy (STS) studies on graphene/hBN heterostructure\cite{29.LeRoy_SDC}, shown as several secondary dips in the tunneling spectra (see Fig.~3(b)). Meanwhile, it is also found that the moir\'e pattern spatially modulates the local density of states (LDOS). Subsequently, magneto-transport measurements confirm the existence of SDCs by observing satellite resistance peaks and the associated Landau fans in the self-similar Hofstadter butterfly spectrum\cite{30.Geim_SDC,31.Kim_SDC,31.2.ZhangGY_NL,32.Ashoori_SDC} (see Figs.~3(c,d)). Further experiments indicate that a set of tertiary Dirac points could even emerge in the heterostructures of graphene/hBN\cite{32.1.ChenGR_TDC}. The observation of fractional quantum Hall effect with $\nu=\pm\frac{5}{3}$ in capacitance measurements suggests the existence of a finite gap opening at the original Dirac points induced by sub-lattice symmetry breaking in this moir\'e superlattice\cite{32.Ashoori_SDC,32.2.WangL_FQHE}.

Although the existence of SDCs has been confirmed, fundamental information on the band structures near SDCs remains unclear due to the lack of momentum resolved information. Theoretical calculations based on different models and parameters lead to different scenarios about the locations of SDCs and whether they are gapped\cite{26.CPark_pp_prl,33.Falko_SDC,34.Brink_SDC,35.MacDonald_SDC,36.Koshino_SDC}. To resolve these issues, experimental electronic structure with both energy and momentum-resolved information is in high demand. The required high-quality and single-domain epitaxial graphene/hBN heterostructures for angle-resolved photoemission spectroscopy (ARPES) measurements are implemented by  remote plasma-enhanced chemical vapor deposition (R-PECVD). Such samples show satellite resistance peaks\cite{37.ZGY_SDC} and similar quantum Hall effect with fan diagram as those samples fabricated by transfer method. Systematic ARPES measurements on epitaxial graphene/hBN heterostructures reveal that the SDCs only occur at only one set of the two superlattice valleys in the moir\'e SBZ (as shown in Fig.~3(e)), indicating the significant role of the asymmetry potential induced by the moir\'e superlattice, which breaks the inversion symmetry of graphene\cite{38.WEY_SDC}. It is further shown that such asymmetric potential leads to gap opening at both the original graphene Dirac cone and SDCs (as shown in Fig.~3(f) and Fig.~1). The ARPES results provide insight into the important role of an inversion-symmetry-breaking perturbation potential in the physics of graphene/hBN vdWH, which is important for determining the actual band parameters for describing the electronic structure engineering of graphene/hBN. In addition, further ARPES measurements on one non-aligned graphene/hBN sample which is fabricated by transfer method, however, do not show obvious trace of the SDCs and gap opening, suggesting a weaker interaction in this kind of transferred sample with a twisting angle\cite{39.WEY_SDC_R}.

\begin{center}
    \includegraphics{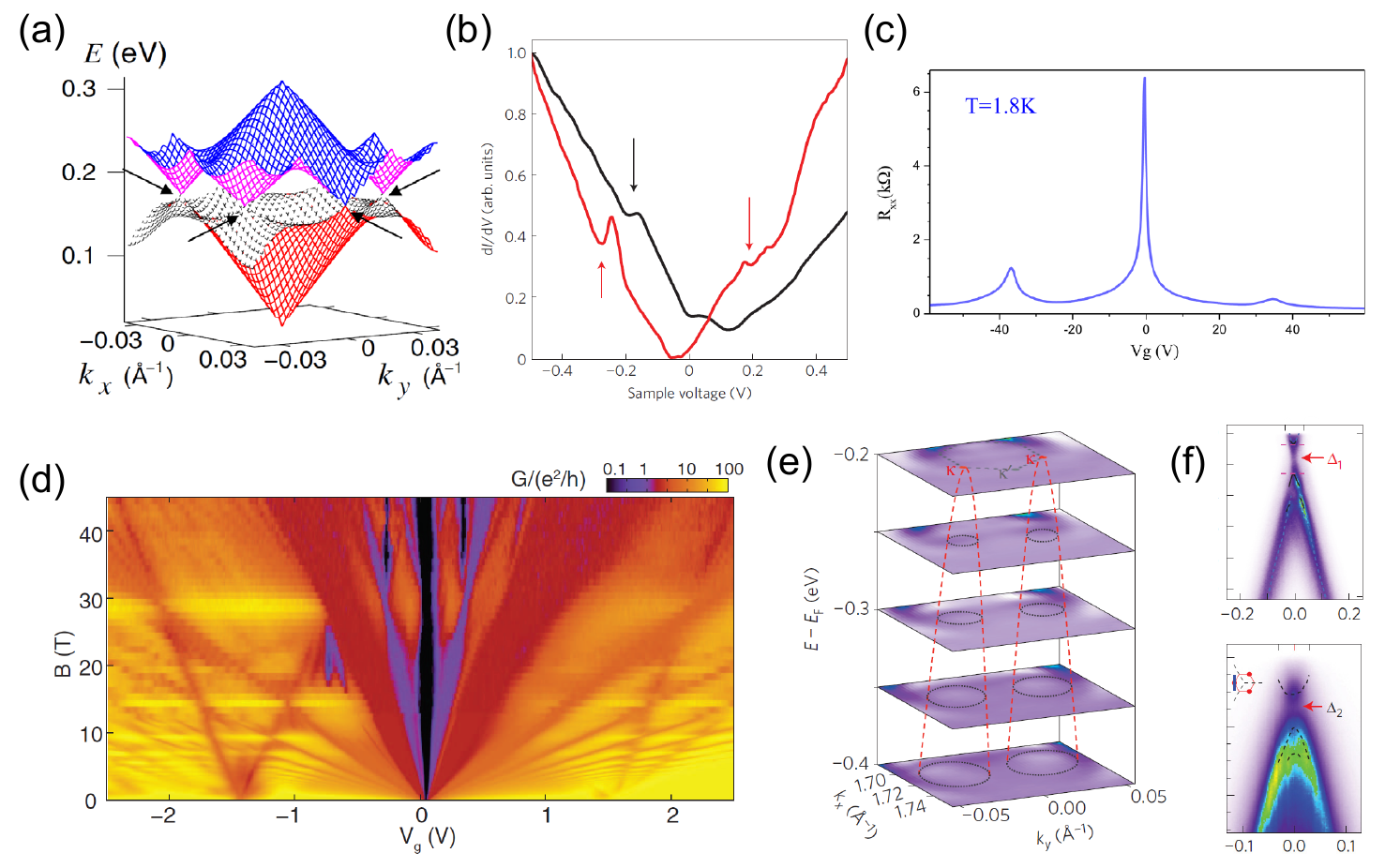}\\[5pt]  
    \parbox[c]{15.0cm}{\footnotesize{\bf Fig.~3.} (color online) (a) Calculated band structures showing the second-generation Dirac cones in graphene/hBN heterostructure (Reprinted with permission from Ref.~\cite{26.CPark_pp_prl}. Copyright 2008 American Physical Society). (b) Secondary dips observed in STS data of graphene/hBN heterostructure (indicated by the arrows) for different moir\'e periods of 9.0 nm (black) and 13.4 nm (red) (Reprinted with permission from Ref.~\cite{29.LeRoy_SDC}. Copyright 2012, Springer Nature). (c) Principal and satellite resistance peaks observed in transport measurements on graphene/hBN heterostructure (Reprinted with permission from Ref.~\cite{31.2.ZhangGY_NL}. Copyright 2016, American Chemical Society). (d) Landau fans in Hofstadter butterfly measured in graphene/hBN heterostructure (Reprinted with permission from Ref.~\cite{32.Ashoori_SDC}. Copyright 2013, American Association for the Advancement of Science). (e) ARPES results for graphene/hBN heterostructure, showing the locations of SDCs (Reprinted with permission from Ref.~\cite{38.WEY_SDC}. Copyright 2016, Springer Nature). (f) ARPES results showing the opened gaps at original Dirac point (top) and second-generation Dirac point (bottom) in graphene/hBN heterostructure (Reprinted with permission from Ref.~\cite{38.WEY_SDC}. Copyright 2016, Springer Nature).}
\end{center}

On the other hand, the observation of insulating states in transport measurements on tBLG suggests the existence of SDCs as well as graphene/hBN heterostructures\cite{40.CaoY_PRL2016,41.Tutuc_PNAS2016}.  ARPES investigation on one tBLG sample with 6\textdegree~twisting angle indicates the sign of a novel type of SDCs locating at the center ($\Gamma$ point) of moir\'e BZ, distinguished from the graphene/hBN sample\cite{42.WEY_Thesis}. The existence of various forms of SDCs in different kinds of vdWHs demonstrate the general ability of moir\'e superlattice potential to tailor the electronic structures of graphene, which could induce more emerging quantum states.

\subsection{van Hove singularity}

Van Hove singularity (VHS) is a point of energy in the electronic structure with diverging density of states (DOS). Saddle point in the band structure, where the effective mass of electron is positive along one momentum direction but negative along its orthogonal direction, is also a VHS. The diverging DOS at VHS is attractive, which could potentially lead to various correlated states like charge density wave (CDW)\cite{43.Shin_NbSe2} and superconductivity\cite{44.Chubukov_chiralSC}.

The VHS in tBLG is firstly confirmed by scanning tunneling microscopic and spectroscopic studies on both the surfaces of highly oriented pyrolytic graphite (HOPG) and CVD-grown samples (transferred from polycrystalline Ni film as growth substrate)\cite{45.Andrei_VHS,62.Andrei_PRL2011}, as shown in Fig.~4(a).  Calculations show that this kind of VHS is resulted from the hybridization gap between two intersecting Dirac cones of each graphene layer or the top two twisted layers in HOPG samples (as illustrated in Fig.~4(b)). Such feature in the band structures is then demonstrated by ARPES measurements on tBLG samples with different twisting angles\cite{46.Eli_VHS,47.CYL_VHS} (see Fig.~4(c)). The VHSs in tBLG always emerge in pairs with one in the occupied state and the other in the unoccupied state, and the dependence of the energy separation between these two VHS points on the twisting angle is subsequently confirmed by experiments and is fitted well with theoretical model\cite{48.Veuillen_VHS,49.HL_PRL2012,50.Crommie_PRB2015} (see Fig.~4(d)). As the twisting angle approaches zero, it is found that the VHS separation in tBLG decreases and they move toward the Fermi energy. For large twisting angles, however, it has been shown that the VHS could disappear\cite{51.HL_PRB2014}. Additionally, the VHS energy separation is shown to be independent of doping for a wide range of twisting angles\cite{52.Veuillen_PRB2015}, but could be strongly affected by strain and curvature in the sample\cite{53.HL_NC2013}. These results indicate a potential way to bring VHS toward the Fermi energy by controlling the twisting angle with the assistance of gate induced doping, strain and so on.  This could lead to possible correlated phenomena, for example, a CDW-like state is indeed observed at the energy of VHS from STS studies\cite{45.Andrei_VHS}.

The VHS in tBLG has significant impact on the optical response. As shown in Fig.~4(e), a large enhancement in the amplitude of the G mode phonon is reported in the Raman spectra of tBLG at specific twisting angles \cite{54.Zettl_PRL2012}. Such an enhancement occurs when the excitation photon energy matches the energy separation between the two VHS points. This suggests that such enhancement can be applied to distinguish tBLG with different twisting angles\cite{55.JPark_NL2012}. Similar enhancement is also observed in photocurrent and photovoltage generation, and photochemical reactivity\cite{56.LZF_NC2016,57.TianJG_AOM2016,58.LZF_NL2015}, opening new routes for graphene-based optoelectronic and photochemical applications.

\begin{center}
    \includegraphics{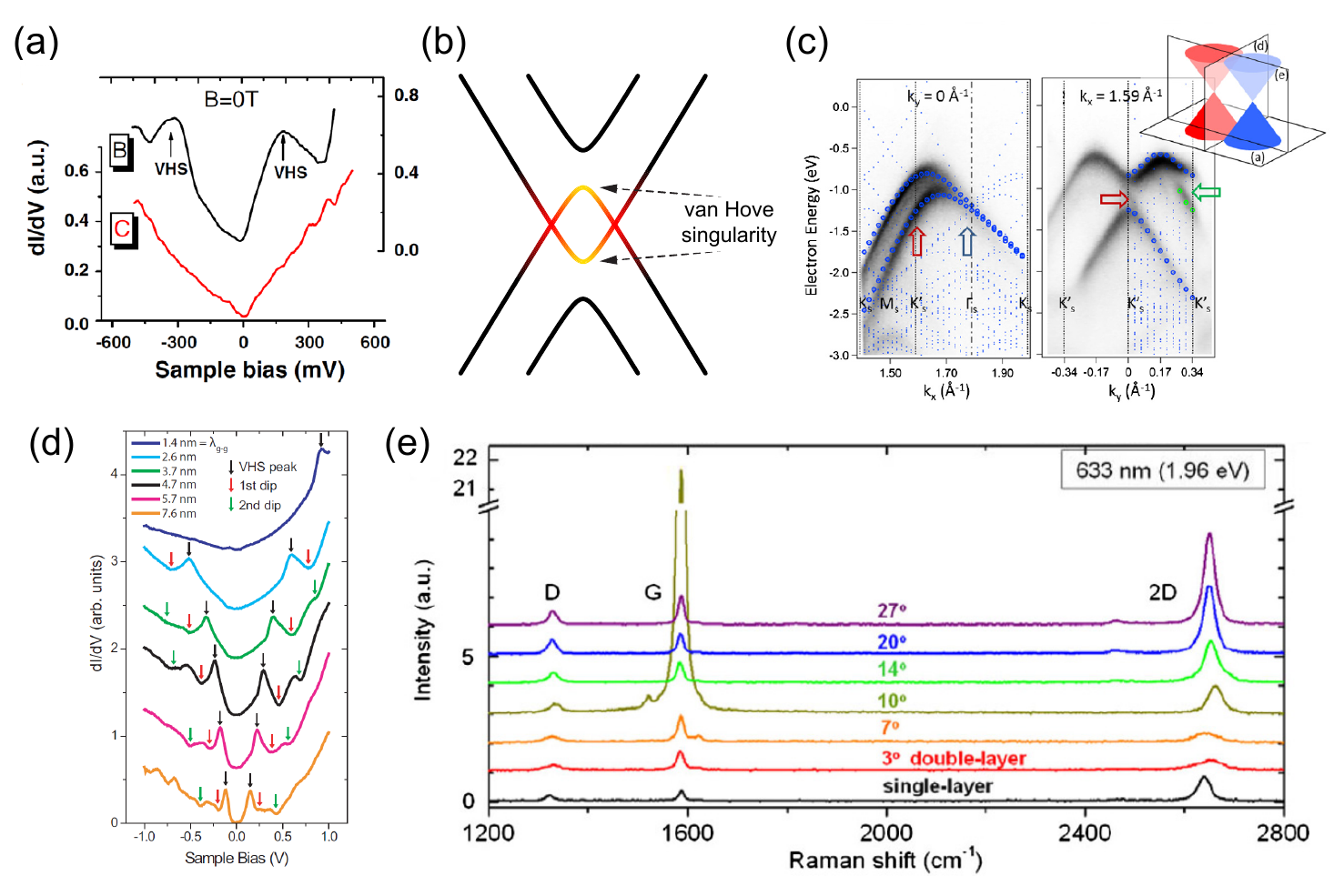}\\[5pt]  
    \parbox[c]{15.0cm}{\footnotesize{\bf Fig.~4.} (color online) (a) STS results on tBLG, indicating the existence of one pair of VHS. The two data curves are taken at different regions: black one (B) in twisting region with moir\'e pattern and red one (C) in non-twisting region without moir\'e pattern (Reprinted with permission from Ref.~\cite{62.Andrei_PRL2011}. Copyright 2011, American Physical Society). (b) Sketch for band structures of tBLG with indication of VHS. (c) Observation of VHS in tBLG from ARPES measurements (Reprinted with permission from Ref.~\cite{46.Eli_VHS}. Copyright 2012 American Physical Society). (d) VHS of tBLG in STS measurements with different twisting angles (Reprinted with permission from Ref.~\cite{50.Crommie_PRB2015}. Copyright 2015, American Physical Society). (e) Raman spectra of tBLG at different twisting angles (Reprinted with permission from Ref.~\cite{54.Zettl_PRL2012}. Copyright 2012, American Physical Society).}
\end{center}

\subsection{Flat bands at magic angle}

\subsubsection{Renormalized Fermi velocity}
The interlayer coupling of tBLG not only leads to VHS at the hybridized gap edge, but also has significant influence on the Fermi velocity of the Dirac cones. Based on a continuum model, J. Lopes dos Santos \textit{et al.} theoretically show that the Fermi velocity $\tilde{v}_{\rm F}$ in tBLG will be renormalized\cite{59.CastroNeto_PRL2007,24.CastroNeto_Continuum} as follows:
\begin{equation}
    \frac{\tilde{v}_{\rm F}}{v_{\rm F}}=1-9(\frac{\tilde{t}_\bot}{\hbar{v_{\rm F}}\Delta{K}})^2 \label{eq4}
\end{equation}
where $v_{\rm F}$ is the original Fermi velocity of pristine monolayer graphene, $\tilde{t}_\bot$ is the Fourier amplitude of the interlayer hopping parameter, and $\Delta{K}$ is the distance between two Dirac points of the two graphene layers. It is clear as depicted in Fig.~5(a) that the Fermi velocity in tBLG dramatically drops as the twisting angle approaches zero, which is consistent with the results of \textit{ab initio} calculations\cite{59.1.TanPH_wlxb2013}. Tight-binding models also reveal the renormalization of Fermi velocity which is suggestive of electron localization\cite{60.Shallcross_PRB2010,61.Magaud_NL2010}. This is experimentally confirmed by STS studies on Landau levels\cite{62.Andrei_PRL2011} (see Fig.~5(a)) and ARPES measurements on the band structures of tBLG system\cite{63.Barinov_SR2015}. 

On the contrary, for some specific samples like tBLG on the carbon face of 4H-SiC, a few studies show the absence of Fermi velocity renormalization, since the $v_{\rm F}$ there is identical to the pristine one\cite{64.Lanzara_PRL2009,65.Vignaud_SR2016,66.Stroscio_Sci2009,67.Stroscio_Nat2010}. This discrepancy is then reconciled by STS studies and shown to be caused by different strength of interlayer coupling in tBLG samples\cite{68.HL_PRB2015}.

\subsubsection{Flat bands and strongly correlated phenomenon}

Obviously, the Fermi velocity renormalization in Eq.~\eqref{eq4} is only valid for small but ``not so small'' twisting angles, since for infinitesimal twisting angles, Eq.~\eqref{eq4} will diverge and fail. When the twisting angle in tBLG structure becomes small enough, moir\'e pattern with a large period then builds up (see Fig.~1), and the corresponding moir\'e mini bands, especially the replica Dirac cones around the original ones, become closer to each other in the reciprocal space. This means that these moir\'e mini bands will impose non-negligible impact on the band structure of tBLG. Subsequent theoretical studies based on tight-binding and low-energy continuum model, with the moir\'e effect taken into account, show that the dependence of Fermi velocity on the twisting angle is not monotonic\cite{69.Barticevic_PRB2010,70.MacDonald_PNAS2011}, and the $v_{\rm F}$ could even vanish at a series of ``magic angles'', leading to the emergence of exotic ``flat bands'' near the charge-neutral Dirac point\cite{70.MacDonald_PNAS2011,71.Koshino_PRB2012,72.FangS_PRB2016} (see Fig.~5(b,c)). The flat bands imply an extremely slow velocity for the quasiparticles and a large DOS in the energy spectrum, which is then observed on HOPG surface by STS measurements\cite{73.HL_PRB2015b}. 

Once the Fermi level is tuned to near the flat bands, correlation-induced phenomena are thus expected to occur. The first experimental breakthrough is the observation of Mott-like insulating states in magic-angle tBLG (with twisting angle of 1.08\textdegree) at half-filling doping level (see Fig.~5(d)), which was firstly realized by Y. Cao \textit{et al.}\cite{74.CaoY_MA1} however with improved fabrication technique and sample quality\cite{75.Dean_twist,76.Tutuc_twist}. In parallel, another major discovery is unconventional superconductivity in magic-angle tBLG when the doping level deviates from half-filling\cite{77.CaoY_MA2} (see Fig.~5(e)), which resembles that of cuprate-based unconventional high-temperature superconductors --- inducing superconductivity by doping a Mott insulator\cite{78.Wen_RMP2006}. Considering the low carrier density in tBLG, the critical temperature $T_{\rm c}$ up to 1.7 K is actually quite ``high'', indicating the contribution from strong electron interaction. The observation of ``strange metal'' behavior in magic-angle tBLG from transport measurements also suggests the resemblance to strongly correlated electron systems\cite{79.Young_MA,80.CaoY_MA3}. Recent studies show that the superconductivity in magic-angle tBLG could be independent from the existence of correlated insulating states\cite{81.Efetov_independentSC,81.Young_independentSC}, raising question on the relation between electron correlation and superconductivity. Further measurements indicate that the magic angles could have a wider range from 0.93\textdegree\cite{81.Lau_MA}~to 1.27\textdegree\cite{82.Dean_MA}, especially under high hydrostatic pressure\cite{83.Kaxiras_PRB2018,84.Jung_MA-pressure}. Magic-angle tBLG is also shown as a platform for hosting orbital ferromagnetism, Chern insulating states and intrinsic quantized anomalous Hall effect\cite{85.Goldhaber_MA,86.Efetov_MA,87.Young_QAHE}, suggesting that more exotic states could emerge in these intriguing tBLG samples.

The correspondence between flat bands and strong electron correlation is not limited to tBLG. Most of the emerging quantum states in magic-angle tBLG, including Mott insulating states, superconductivity, orbital ferromagnetism and Chern insulating states, are also shown to appear in another material system of ABC-stacked trilayer graphene moir\'e superlattice (placed on hBN as the substrate)\cite{87.CGR_MA1,88.CGR_MA2,89.CGR_MA3}. At the same time, correlated insulating states and spin-polarized ground states are theoretically predicted and experimentally demonstrated to exist in one more complicated heterostructure --- twisted double bilayer graphene (TDBG)\cite{90.Vishwanath_TDBG,91.Tutuc_TDBG,92.CaoY_TDBG,93.ZhangGY_TDBG,94.Yankowitz_TDBG,95.Kim_TDBG}, showing remarkable tunability by controlling twisting angle and external electric displacement field. All these novel correlated phenomena can be attributed to similar flat bands induced by the moir\'e periodic potential in these kinds of heterostructures\cite{87.CGR_MA1,90.Vishwanath_TDBG,96.Jung_TDBG,97.Koshino_TDBG}, demonstrating the essential role of flat bands in producing strong electron correlation.

The signatures of flat bands and the induced correlated phenomena in magic-angle tBLG are also revealed in recent spectroscopic studies. In STS measurements, characteristic features of Mott insulators, such as peak broadening and gap opening, are observed in magic-angle tBLG under the condition of half-filling for one flat band\cite{98.Yazdani_MA,99.Pasupathy_MA,100.Nadj_MA}. It has been suggested that in order to explain the origin of these features in the tunneling spectrum, strong electron correlation must be taken into account. The rotational symmetry of the electronic states in magic-angle tBLG is found to be broken in the correlated insulating regime, suggesting possible existence of nematic charge order\cite{99.Pasupathy_MA,100.Nadj_MA,101.Andrei_MA}. Moreover, recent STS studies further show novel cascade of electronic phase transitions in tBLG at integer fillings\cite{102.Yazidani_cascade}, which is also observed by electronic compressibility measurements\cite{102.Ilani_cascade}, revealing the strong-correlation-induced breaking of spin-valley- or flavor-symmetry. In recent ARPES studies, the flat bands in magic-angle tBLG have been directly visualized by using Nano-ARPES technique\cite{102.WangF_flatband,103.Baumberger_flatband}. Spectral function simulations qualitatively reproduce the observed band structure and opening of hybridization gaps, thus experimentally confirming the existence of weakly dispersive flat bands in the momentum space (see Fig.~5(f,g). This is a first important step in directly resolving the flat bands. Future Nano-ARPES or ARPES studies on more uniform samples with precisely controlled twisting angle will be important to reveal more detailed band structure engineering and to check how sensitive such flat bands depend on the twisting angle. Moreover, ARPES measurements with {\it in situ} gating below the superconducting transition temperature is critical to provide information on the symmetry of the superconducting pairing gap.

\begin{center}
    \includegraphics{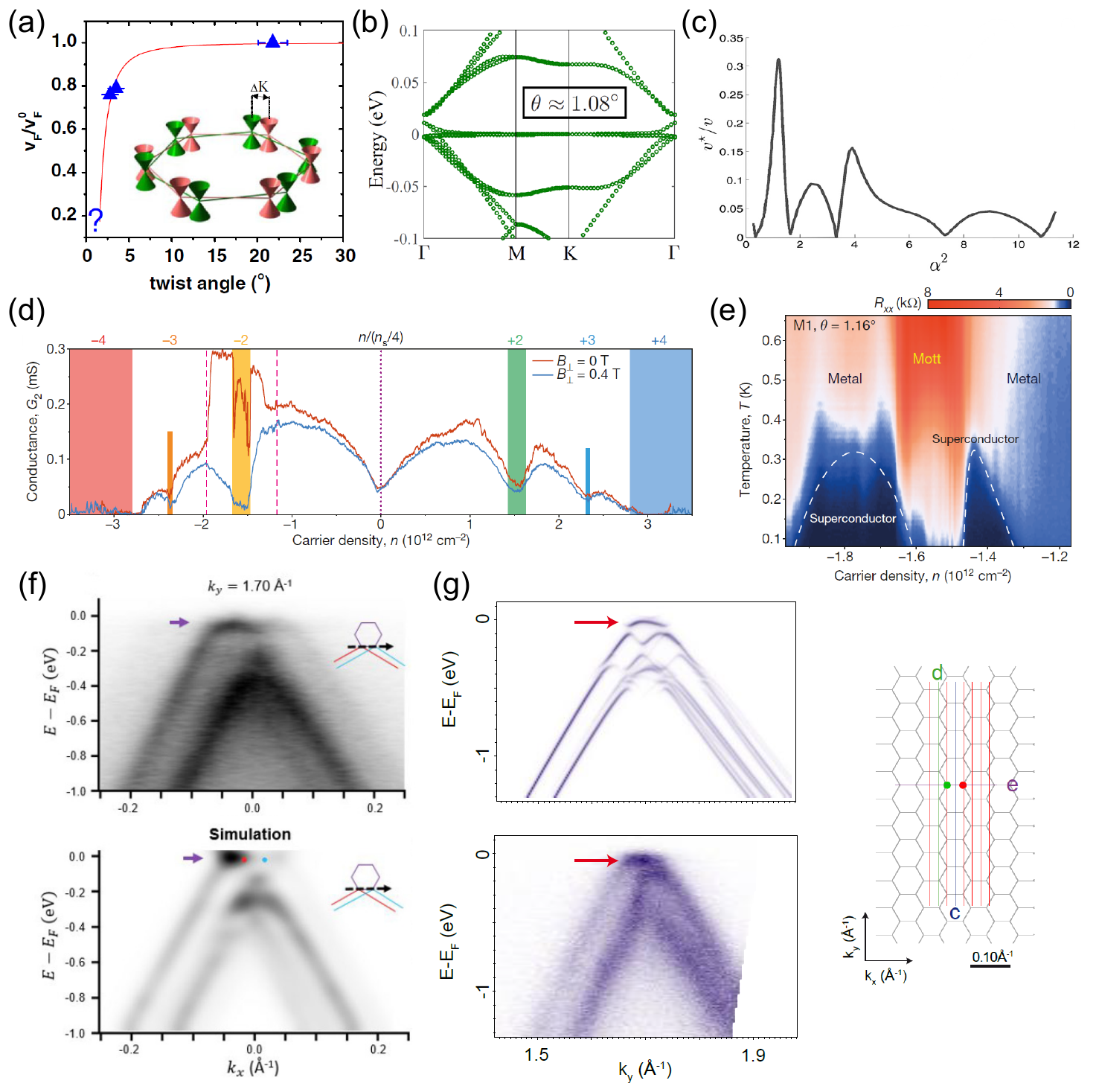}\\[5pt]  
    \parbox[c]{15.0cm}{\footnotesize{\bf Fig.~5.} (color online) (a) Measured Fermi velocity by STS from tBLG with different twisting angles. The curve is plotted according to the calculations (Reprinted with permission from Ref.~\cite{62.Andrei_PRL2011}. Copyright 2011, American Physical Society). (b) Calculated band structures of tBLG at magic angle, showing almost flat bands (Reprinted with permission from Ref.~\cite{72.FangS_PRB2016}. Copyright 2016, American Physical Society). (c) Twisting-angle-dependence of Fermi velocity for tBLG with extremely small angles, showing several magic angles with zero Fermi velocity (Reprinted with permission from Ref.~\cite{70.MacDonald_PNAS2011}. Copyright 2011, National Academy of Sciences). (d) Observation of correlated insulating states at half-filling for the flat band of magic-angle tBLG (Reprinted with permission from Ref.~\cite{77.CaoY_MA2}. Copyright 2018, Springer Nature). (e) Superconductivity observed in magic-angle tBLG (Reprinted with permission from Ref.~\cite{77.CaoY_MA2}. Copyright 2018, Springer Nature). (f,g) ARPES results for the flat bands in magic-angle tBLG (top panel in (f) and left bottom panel in (g) ), and comparison with spectral function simulation (bottom panel in (f) and left top panel in (g)). Insets in (f) and right panel in (g) indicate the cut lines of the spectra in k-space. (Reprinted from Ref.~\cite{102.WangF_flatband,103.Baumberger_flatband}).}
\end{center}


\section{Band structure engineering in incommensurate superlattices}

Although the commensurate superlattices possess rich emerging phenomena as discussed above, incommensurate ones are more common among all configurations of vdWHs. Taking tBLG as an example, Eq.~\ref{eq2} and Fig.~2 indicate that commensurate superlattices are a series of twisting angles with discrete values, while the rest portion of twisting angles are incommensurate superlattices. It is hence more likely to have incommensurate structure for a randomly stacked tBLG sample, especially for those with twisting angle near 30\textdegree. A generic question to ask is what happens to the band structure of such incommensurate superlattice? Will the electronic structure still be affected by the incommensurate superlattice potential?

Intuitively, the broken translational symmetry in incommensurate tBLG lowers the coherence of electron motion between the layers, which makes the two graphene layers in incommensurate tBLG behave as two independent graphene monolayers. From the view of electronic structure, the absence of long-ranged periodicity leads to the failure of Bloch theory in such incommensurate vdWHs, which have attracted much less attention in the past. However, recent experimental progress not only challenges this intuition, but also reveals newly emerging electronic structures and generic electron scattering mechanism in the incommensurate vdWHs, which is demonstrated in a 30\textdegree-tBLG with quasicrystalline symmetry\cite{106.Yao_30tBLG,107.Ahn_30tBLG}.

\subsection{Restored Dirac cone by twisting}

One significant difference of tBLG from the aligned Bernal-stacking bilayer graphene (with 0\textdegree~twisting angle) is the persistence of Dirac points for its $\pi$ bands\cite{59.CastroNeto_PRL2007}. Although it is suggested theoretically that the massless Dirac point in commensurate tBLG could be modified to finite gap or effective mass due to the breaking of so-called ``sublattice-exchange symmetry''\cite{104.Mele_PRB2010}, the band structures of incommensurate tBLG are believed to remain similar to the pristine ones, leading to the viewpoint of interlayer decoupling. This viewpoint for incommensurate tBLG is supported by ARPES measurements on epitaxial multi-layer graphene grown on the carbon face of SiC\cite{64.Lanzara_PRL2009}, where the interlayer interaction is suggested to be weakened by the twisting. Similar view on the negligible interlayer conduction has also been suggested from transport measurements\cite{105.twist_decoupling}. Recent experimental progress in 30\textdegree-tBLG\cite{106.Yao_30tBLG,107.Ahn_30tBLG} suggested that interlayer coupling could be significant, leading to tailored electronic band structure as well.

\subsection{30\textdegree-tBLG as a two-dimensional quasicrystal}

Although commensurate vdWHs with long-ranged periodicity, which exhibit a rich variety of intriguing phenomena, have been discovered in the family of 2D materials, exotic 2D quasicrystal with only rotational symmetry but no translational symmetry has not been realized until tBLG with 30\textdegree~twisting angle becomes true experimentally. As shown in Fig.~1, the incommensurate structure of 30\textdegree-tBLG lacks long-ranged periodicity but shows 12-fold rotational symmetry. Such a superlattice pattern resembles that of dodecagonal quasicrystal, which can be fully filled by ingredients with two simple shapes (square and regular triangle)\cite{106.Yao_30tBLG,107.Ahn_30tBLG,108.Superlubricity}. More interestingly in mathematics, this quasicrystalline pattern can be viewed as the projection of a 12-dimensional hypercubic lattice or 4-dimensional hyperhexagonal lattice onto the 2D plane\cite{109.QCmath1,110.QCmath2}, and has fractal-like structure with dilatation of $2+\sqrt{3}$\cite{108.Superlubricity}. 

Such a quasicrystal has been proposed theoretically to host exotic 4D integer quantum Hall effect\cite{111.4D-QHE}. More elaborate theoretical studies on the electronic structures predict localized behaviors for the electrons in this non-periodic heterostructure\cite{112.Koshino_30tBLG,113.ParkM_30tBLG}. Experimentally, the 30\textdegree-tBLG samples have been successfully grown on surface of different substrates, such as Pt(111)\cite{106.Yao_30tBLG} and the Si face of 4H-SiC\cite{107.Ahn_30tBLG}, where the features of quasicrystal have been revealed by low energy electron diffraction (LEED) and transmission electron microscopy (TEM) measurements (as shown in Fig.~6(a)). More recently, 30\textdegree-tBLG has also been realized on Cu(111) substrate\cite{114.HL_30tBLG,115.PengHL_30tBLG}, where STM and TEM studies visualize the 12-fold dodecagonal quasicrystalline order.


\begin{center}
    \includegraphics{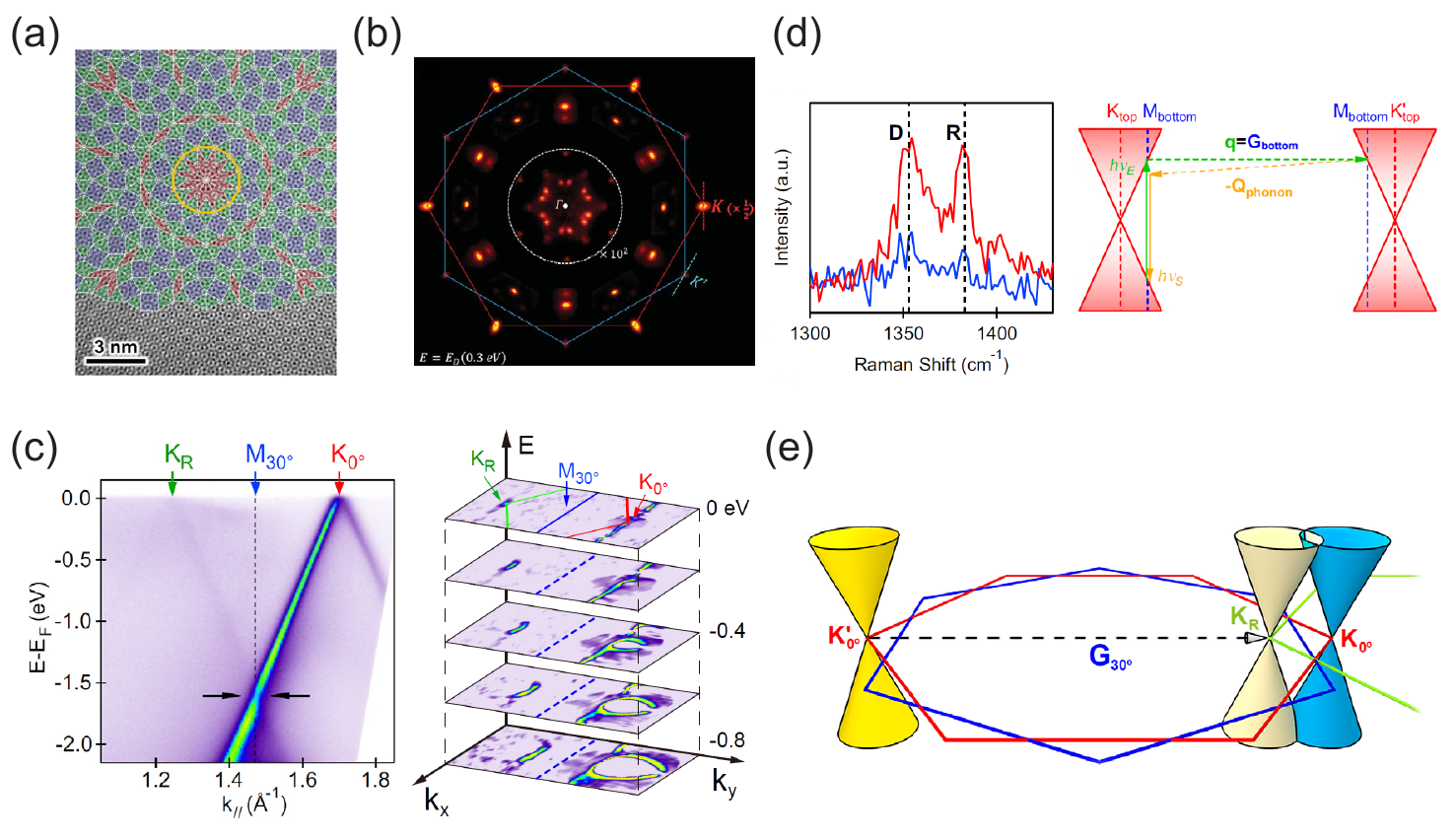}\\[5pt]  
    \parbox[c]{15.0cm}{\footnotesize{\bf Fig.~6.} (color online) (a) TEM image of the quasicrystalline pattern in 30\textdegree-tBLG (Reprinted with permission from Ref.~\cite{107.Ahn_30tBLG}. Copyright 2018, American Association for the Advancement of Science). (b) Constant energy map in ARPES results of 30\textdegree-tBLG at the level of Dirac points (Reprinted with permission from Ref.~\cite{107.Ahn_30tBLG}. Copyright 2018, American Association for the Advancement of Science). (c) Left: ARPES cut along $\Gamma$-K direction of one layer in 30\textdegree-tBLG, showing the mirrored replica Dirac cone and the opened hybridization gap. Right: constant energy maps showing mirrored relation in 2D $k$-space (Reprinted with permission from Ref.~\cite{106.Yao_30tBLG}. Copyright 2018, National Academy of Sciences). (d) Left: observation of additional R peak in Raman spectrum of 30\textdegree-tBLG. Right: process of double-resonant Raman scattering. (Reprinted with permission from Ref.~\cite{106.Yao_30tBLG}. Copyright 2018, National Academy of Sciences). (e) Scattering mechanism for electrons in quasicrystalline 30\textdegree-tBLG. (Reprinted with permission from Ref.~\cite{106.Yao_30tBLG}. Copyright 2018, National Academy of Sciences).}
\end{center}

\subsection{Mirrored Dirac cone and interlayer coupling in 30\textdegree~twisted bilayer graphene}

The evidence for interlayer coupling in epitaxial 30\textdegree-tBLG on Pt(111) and SiC substrates has been revealed by ARPES measurements. Besides the original Dirac cones from the two graphene layers, additional replica Dirac cones are also observed in the Fermi surface maps on both samples grown on Pt(111) and 4H-SiC substrate\cite{106.Yao_30tBLG,107.Ahn_30tBLG} (see in Fig.~6(b,c)). The location and shape of these replica Dirac cones of one graphene layer show particular mirror symmetry with respect to the original ones with the BZ boundary of the other layer as the mirror plane (see Fig.~6(c)), and therefore they are called mirrored Dirac cones. At the intersection between the original and mirrored Dirac cones (also the mirror symmetry center and BZ boundary), hybridized energy gap is observed, suggesting the existence of interlayer coupling in 30\textdegree-tBLG. Such features of mirrored Dirac cones and gaps could be well reproduced by calculations\cite{116.YuanSJ_30tBLG}. The origin of mirrored Dirac cones is then attributed to one particular electron scattering mechanism in which interlayer tunneling and Umklapp scattering are both involved\cite{106.Yao_30tBLG,107.Ahn_30tBLG}. Thus, as shown in Fig.~6(e), the emergence of mirrored Dirac cones is a result of electrons in one graphene layer scattered by one reciprocal lattice vector of the other graphene layer (obeying ${\bm k}^{\prime}_1={\bm k}_1+{\bm G}_2$), showing significant interlayer coupling. The above scattering mechanism is also supported by observing high-order scattered Dirac cones in ARPES results\cite{107.Ahn_30tBLG}. Moreover, Raman measurements further confirm the proposed scattering mechanism by observing a novel Raman R peak from double-resonance scattering\cite{106.Yao_30tBLG,117.Coletti_30tBLG} (see Fig.~6(d)). Indeed, such R peak is quite sensitive to the twisting angles, hence it suggests that the R peak can be used to characterize the twisting angle of tBLG. With this scattering mechanism in which electrons are scattered in the momentum space from state ${\bm k}_1$ to state ${\bm k}^{\prime}_1={\bm k}_1+{\bm G}_2$, the interlayer coupling in incommensurate tBLG is definitively revealed by combining ARPES and Raman measurements. It is worth noting that such scattering mechanism does not require a long-ranged translational symmetry, and thus could generally apply to vdWHs no matter whether the superlattices are commensurate or incommensurate, which is valuable in designing future 2D electronic devices.

More recently, some up-to-date results on 30\textdegree-tBLG have been obtained by spectroscopic studies. Ultrafast response of the electrons in 30\textdegree-tBLG is studied by time-resolve ARPES, revealing unbalanced electron distribution between two graphene layers and between original and mirrored Dirac cones due to the influence from substrate\cite{118.Matsuda_30tBLG}. STS measurements on 30\textdegree-tBLG, especially for those grown or transferred on Cu(111), both indicate the linear dispersions as usual, which concludes from model fitting for the curves of magnetic-field-dependent Landau level peak positions\cite{114.HL_30tBLG} and gating-dependent Dirac point energy\cite{115.PengHL_30tBLG} respectively. The fitted Fermi velocity is similar to the value $\sim 1\times10^{6}$ m/s in pristine graphene, which is consistent with ARPES results and suggests that 30\textdegree-tBLG is still one system with relativistic Dirac fermions in low energy regime. The same conclusion is also obtained by later magnetotransport studies\cite{117.Coletti_30tBLG}. In addition, the modification in electronic structures at high energy regime is revealed in STS measurements by observing a kink in the DOS which is attributed to the hybridizing gap between the original and mirrored Dirac cones as shown in ARPES results\cite{114.HL_30tBLG}. However, the limitation on momentum resolution makes those studies inadequate for investigating the electronic structure in detail such as the existence of mirrored Dirac cones from the DOS.

\section{Conclusions and Outlook}

Van der Waals heterostructures and twisted bilayer materials, especially the graphene/hBN and tBLG as two model systems, have provided a rich playground for studying various emerging quantum states with rich physics. As a tuning knob for electronic properties, twisting plays a central role not only for discovering novel emerging phenomenon in the twisted heterostructures, but also for providing reliable methods to engineer band structures of materials in future device applications called twistronics\cite{118.2.twistronics1,118.3.twistronics2}. Therefore twisting is believed to become an effective and regular method soon for exploring exotic electronic properties in different kinds of vdWHs.

In addition to what has been demonstrated, the field for twisted vdWHs is still in rapid development. For tBLG, a few novel quantum phenomenon, including neutrino-like states\cite{119.neutrino} and topological crystalline insulator states\cite{120.TCI} in commensurate tBLG, high-order topological insulator in 30\textdegree-tBLG\cite{121.HOTI}, and so on, have been predicted by theory and await experimental realization. The superconductivity in tBLG could even be stabilized by monolayer WSe$_2$ as a substrate without the insulating states\cite{122.MA+WSe2}. Furthermore, researchers are looking for similar or even more exotic electronic properties in other 2D materials beyond graphene, such as TMDC\cite{123.tTMDC1,124.tTMDC2,125.tTMDC3}, hBN\cite{126.tBN}, black phosphorus\cite{127.tBP} and GeSe\cite{128.tGeSe}. Some experimental progress have been achieved recently, such as the observations of moir\'e flat bands\cite{129.LeRoy_tTMD}, flat-band-induced correlated electronic phases\cite{129.WangL_tTMD}, and the moir\'e-trapped valley excitons\cite{130.WangF_moireX,131.LiXQ_moireX,132.XuXD_moireX,133.Tartakovskii_moireX,134.Imamoglu_moireX,135.Gerardot_moireX,136.ParkH_moireX,137.ZhuXY_moireX} in many kinds of twisted TMDC hetero-bilayers. At present stage, one of the challenges in practice is how to prepare high quality samples with accurately controlled alignment. With further improvement in sample fabrication techniques, the introduction of twisting will definitely expand the phase diagram of 2D materials and heterostructures, and could lead to emerging field of twistronics.


\end{document}